\documentclass[draftclsnofoot,onecolumn]{IEEEtran}
\pdfoutput=1
\usepackage{cite}
\usepackage{mathtools}
\usepackage{bbm}
\usepackage{float}
\usepackage[colorlinks,linkcolor=red,anchorcolor=red,citecolor=red]{hyperref}
\usepackage[pdftex]{graphicx}%
\usepackage[justification=centering]{caption}
\graphicspath{{./graphics/}}
\usepackage{subfigure}
\usepackage{paralist}
\usepackage{bm}
\usepackage{amsfonts}
\usepackage{multirow}
\usepackage{amssymb}
\usepackage{color}
\usepackage{array}
\makeatletter
\newif\if@restonecol
\makeatother

\usepackage[linesnumbered,ruled,vlined]{algorithm2e}
\usepackage{algpseudocode}
\usepackage{fixltx2e}
\usepackage{stfloats}

\usepackage{amsthm}

\begin{document}
	\title{Joint Resource Optimization for IRS-Assisted mmWave MIMO under QoS Constraints }  
	
	\author{Qingfeng Ding,~\IEEEmembership{Member,~IEEE,} Xinpeng Gao, Zexiang Wu 	
		\thanks{This work was supported by the National Natural Science Foundation of China (No. 61961018), the Jiangxi Province Foundation for Distinguished Young Scholar (No.20192BCB23013), the Jiangxi Province Natural Science Foundation of China (20192ACB21003).}	
		\thanks{Q. Ding, X. Gao and Z. Wu are with the School of Electrical and Automation Engineering, East China Jiaotong University, Nanchang 330013, China (corresponding author to provide e-mail: brandy724@sina.com).}}
	
	
	\maketitle
	
	\begin{abstract}
		This letter focuses on the non-convex joint optimization with a dynamic resource of multi-user for an intelligent reflecting surface-enhanced mmWave system, where all users are concentrated on the unique cluster beam.   Firstly, the objective function of the above non-linear problem is converted into a quadratic programming form under the quality of service constraints. Further, a multi-blocks alternating optimization framework with dynamic power allocation is proposed to obtain the maximum sum-rate, where the relaxed ADMM algorithm is adopted to tackle the optimal full-digital precoder and the corresponding passive reflecting matrix is obtained by the gradient-projection. The numerical results verify that beam optimization should be emphasized in high SINR, but joint dynamic resource allocation can further improve system performance even if the hardware dimensions reaches the limit.
	\end{abstract}
	
	
	\IEEEpeerreviewmaketitle
	
	\section{INTRODUCTION}
	\label{introduction}
	
	\IEEEPARstart{M}illimeter wave (mmWave) massive multi-input multi-output (MIMO) technology, relying on abundant spectrum resources and high multiplexing gain, can provide superior spectral efficiency (SE) and reliable low-latency service for ultra-high rate communications \cite{MMMC2018SA}. However, for communication networks with  active multi-user, the quality of service (QoS) keeps decreasing due to severe attenuation by the blocked of direct links. It motivates the research in intelligent reflecting surface (IRS) which enhances the mmWave MIMO technology by a large number of passive reflective elements with relatively low cost, to rebuilding the additional links \cite{Towa2020WQ}. 
	
	Naturally, a new performance bottleneck is taken from the novel hardware structure which further reduces the accuracy and punctuality that system demand \cite{MMCR2020CP}. Recently, researchers hope to improve the performance limit of IRS-assisted wireless networks by the multi-objective joint optimization \cite{Prac2020WC,IRSCE2020BZ,BOIRS2020BZ}. Among them, a theoretical framework of reflection coefficients was presented in \cite{Prac2020WC} which lays a foundation for further jointly optimizing the transmitter and IRS. The authors in \cite{IRSCE2020BZ} optimized the reflection parameters for an IRS-enhanced system with frequency-selective channels. It was shown that based on practical transmission protocol, a larger area IRS reflection unit can achieve comparable system performance but the training overhead for optimization will increase significantly. The work in \cite{BOIRS2020BZ} further improved the system performance of the IRS-assisted MIMO architecture by adopting a new sum-path-gain maximization criterion based on the alternating direction method of multipliers (ADMM). However, for the communication networks with multiple active users, the performance improvement is hampered by the resource competition and interference among different terminals  \cite{IRSAM2020GZ,STfIRS2020JQ}.

	Motivated by the above, many existing works have looked into the joint optimization of different multi-user structures. To maximize the energy efficiency of the multi-user system, \cite{2020arXiv200500269Y} reconstructed the optimization problem of the joint switch state and reflection coefficient, which was solved by the low complexity successive convex approximation algorithm. In \cite{ASDoIRS2020ZD}, to ensure the communication quality of cell-edge users, the IRS was adopted to extend the beam coverage in the non-orthogonal multiple access (NOMA) system. To verify the effectiveness of the IRS in the NOMA, \cite{OPMfIRS2020HW} proposed a joint optimization algorithm based on the correlation of the transmit power consumption and the phase shifts of the IRS.  Besides,  \cite{HPBMM2019DL} confirmed that the power and information transmission requirements of multi-user in a point-to-point system can be guaranteed by the joint optimization of the power allocation (PA) and the beamforming at the base station (BS). However, due to the additional modulus constraints, the joint optimization of multiple variables is still a challenging problem in the IRS-assisted system with intensive active users.

	This letter is on the joint optimization in the IRS-assisted massive MIMO system with mmWave channel, multi-user interference, and the QoS constraints with power limitations. Our goal is to maximize the sum-SE by jointly optimizing the full-digital precoder at the BS with dynamic PA and the passive beamforming at the IRS.  Therefore, we reconstruct the objective function to decouple the joint optimization variables, based on interference cancellation by superposition coding and inverse matrix transformation. Among them,  an alternating optimization framework is proposed for the quadratic programming problem to obtain the maximization sum-gain, where the relaxed ADMM algorithm is adopted to tackle the optimal full-digital precoder at the BS under QoS constraints. Moreover, the suboptimal solution of phase shifters in the IRS is guaranteed by the gradient-projection method and an analytical solution of the multi-user dynamic PA  is provided for the target variables. Numerical simulations validate the effectiveness of our proposed joint optimization approach.
	
	\section{SYSTEM DESCRIPTION}
	\label{sec:SYSTEM DESCRIPTION AND PROBLEM FORMULATION}
	
	Consider the downlink of an IRS-assisted single-cell mmWave communication system, where the IRS with the uniform rectangular array is deployed in the vicinity of $K$ single-antenna users. The BS is equipped with a uniform linear array of $N_\mathrm{t}$ antennas to support the communication for multi-users. Due to the frequency fading characteristics and insurmountable barriers, the direct path channel between the BS and users is blocked.
	
	\subsection{Channel model}
	It is assumed that the IRS consists of $N_\mathrm{r}$ passive reflecting elements which adjust the IRS pattern for desired signal reflection, and the $K$ users are grouped into $1$ cluster.  Among them, the widely used Saleh-Valenzuela model is adopted for the BS-IRS and the IRS-User mmWave links \cite{HPD2020CP}. Let $N_\mathrm{r} = N_\mathrm{rx}\times N_\mathrm{ry}$, where $N_\mathrm{rx}$ and $N_\mathrm{ry}$ as the number of passive reflecting elements of the IRS on horizontal and vertical, respectively. Assume that there are $L_\mathrm{BI}$ propagation paths for the BS-IRS link, $\alpha _{n_\mathrm{BI}}^{{\mathrm{AoA}}}(\phi _{n_\mathrm{BI}}^{{\mathrm{AoA}}})$ and $\alpha _{n_\mathrm{BI}}^{{\mathrm{AoD}}}$ denote the azimuth (elevation) angle-of-arrival (AoA) and angle-of-departure (AoD) respectively of the $n$th propagation path in the BS-IRS link , where $\alpha _{n_\mathrm{BI}}^{{\mathrm{AoA}}}$, $\phi _{n_\mathrm{BI}}^{{\mathrm{AoA}}}$ and $\alpha _{n_\mathrm{BI}}^{{\mathrm{AoD}}}\in\left[ { - \frac{\pi }{2},\frac{\pi }{2}} \right] $. The channel matrix between the BS and IRS is modeled as
	\begin{equation}
	\mathbf{H}=\sqrt{\frac{N_{\mathrm{t}} N_\mathrm{r}}{L_\mathrm{BI}}} \sum_{n=1}^{L_\mathrm{BI}} \mathfrak{p}^{\mathrm{BI}}_{n} \mathfrak{a}_{n}\left(\alpha_{n_\mathrm{BI}}^{\mathrm{AoA}}, \phi_{n_\mathrm{BI}}^{\mathrm{AoA}}\right) \mathfrak{b}_{n}^{H}\left(\alpha_{n_\mathrm{BI}}^{\mathrm{AoD}}\right),
	\end{equation}
	where $\mathfrak{p}^{\mathrm{BI}}_{n}$ is the complex channel gain of $n$th path in the BS-IRS link, $\mathfrak{a}_n$ and $\mathfrak{b}_n$ are the receive and transmit array response vectors of the $n$th propagation path and similar as the setting in \cite{2019arXiv190810734W} and therefore omitted for brevity. Similarly, the channel vector between the IRS and $kth$ users $\mathbf{h}_k$ is modeled as
	\begin{equation}
	\mathbf{h}_{k}=\sqrt{\frac{N_{\mathrm{r}}}{L_\mathrm{IU}}} \sum_{n=1}^{L_{\mathrm{IU}}} \mathfrak{p}_{k, n}^{\mathrm{IU}} \mathfrak{a}_{n}\left(\alpha_{k, n_\mathrm{IU}}^{\mathrm{AoD}}, \phi_{k, n_\mathrm{IU}}^{\mathrm{AoD}}\right),
	\end{equation}
	where $L_\mathrm{IU}^k$ is the number of propagation paths of the $k$th IRS-User link, $\mathfrak{p}_{k, n}^{\mathrm{IU}}$ is the complex channel gain of $n$th path in the $k$th IRS-User link, $\alpha_{k, n_\mathrm{IU}}^{\mathrm{AoD}}$ and $\phi_{k, n_\mathrm{IU}}^{\mathrm{AoD}}$ are the azimuth and elevation AoD, respectively.

	\subsection{Downlink signal transmission process}
	 In the communications, the BS send its data message vector $\mathbf{s}_k$ for the $k$th user with normalized power $\mathrm{E}(\mathbf{s}_\mathrm{k}\mathbf{s}_\mathrm{k}^H)= 1$ to the IRS, where $k \in [1,2,\cdots,K]$. The received signal at the IRS is first shifted by a diagonal reflection matrix ${\mathbf{\Theta}}{\mathrm{ = diag(}}\beta \theta _1,\beta \theta _2{\mathrm{,}} \cdots ,\beta \theta_{N_\mathrm{r}}{\mathrm{)}}\in\mathbb{C}^{N_\mathrm{r}\times N_\mathrm{r}}$ and then reflected to users, where $\theta_i = {e^{j{\varphi_i}}} $ and the phase shifters of each element $\varphi_i \in [-\pi, \pi), i = 1,2,\cdots,N_\mathrm{r}$. To ensure correct decoding, we assign sequence labels  by the ascending order of channel strength, i.e.,  $\|\mathbf{G}_1\|_2 \geq \|\mathbf{G}_2\|_2 \geq \cdots\geq\|\mathbf{G}_k\|_2\geq\cdots \geq\|\mathbf{G}_j\|_2\geq \cdots \geq\|\mathbf{G}_K\|_2 $, where $\mathbf{G}_k = \mathbf{h}_k\Theta\mathbf{H}\mathbf{F}$ and  $\mathbf{F} \in\mathbb{C}^{N_\mathrm{t}\times1}$ denotes the full-digital precoding matrix at BS. Note that, thanks to the perfect successive interference cancellation and the difference of channel quality, the $k$th user can detect and remove the $j$th user’s signals interference for all of $1 \leq k \le j \leq K$ in a successive manner  \cite{STvJP2019ZN}. Thus, the remaining receiver signal vector at the $k$th user can be modeled as
	\begin{equation}
	\label{rewritten signal model}
	{{\mathbf{y}}_k} ={\mathbf{G}_k\sqrt {{p_k}} {{\mathbf{s}}_k}}+ {\mathbf{G}_k\sum\limits_{i = 1}^{k - 1} {\sqrt 	{{p_i}} {{\mathbf{s}}_i}}}+ {{{\mathbf{n}}_k}}.
	\end{equation}
	where the first item denotes the desired signal for users and  $p_k$ is the transmitted power for the $k$ user satisfies $\sum_{k = 1}^K {{p_k}}  \le P$, the last item  ${{\mathbf{n}}_k}$ denotes the noise vector whose entries are dependent and identically distributed as $\mathcal{C} \mathcal{N}\left(\mathbf{0}, \sigma^{2} \right)$, the remaining items represent intra-cluster interference.
	
	\subsection{Analysis of achievable rate}
	
	To obtain the optimal design and the corresponding performance upper bound, we assuming that perfect channel state information of all channels is available and the BS with the maximum transmitted power $P$. According to \eqref{rewritten signal model}, the achievable rate of the proposed scheme  can be presented as
	\begin{equation}
	\label{RSUM}
	R_\mathrm{sum} = \sum_{k=1}^{K}\log_{2}\left(1+\gamma_k\right),
	\end{equation}
	where $\gamma_k=\frac{\|\mathbf{h}_{k}\Theta\mathbf{H}\mathbf{F}\|_\mathrm{F}^2 p_k}{\|\mathbf{h}_{k}\Theta\mathbf{H}\mathbf{F}\|_\mathrm{F}^2\sum_{i=1}^{k-1}p_i+\sigma_k^2}$ which can be improved by carefully designing the angle of the IRS $\mathbf{\Theta}$ , precoding matrix $\mathbf{F}$ and user transmission power $\{p_k\}_{k=1}^K$. Without loss of generality, define $\mathbf{G}_\mathrm{max} = \mathop{\mathrm{argmax}}\limits_{k =1,2,\cdots,K}{\{\mathbf{G}_k\}}$ denotes the equivalent matrix of the destination with the strongest channel response in the beam and considering achievable rate characteristics of users. When using high signal-to-interference-plus-noise-ratio (SINR) approximation and ignored constant terms, the achievable rate can be transformed as	$\hat{R}_\mathrm{sum} \approx \log_{2}(\|\mathbf{G}_\mathrm{max}\|_\mathrm{F}^2/\sigma^2)+\log_{2}(\sum_{i=1}^{K}p_i) =\log_{2}({P\|\mathbf{G}_\mathrm{max}\|_\mathrm{F}^2/\sigma^2})$. It is not difficult to observe that since each user receives a signal with sufficient transmission power, the performance loss of sum rate $\mathbf{R}_\mathrm{sum}$ caused by different power distribution factors $p_k$ and channel strength $\mathbf{h}_k$  can be compensated but still effective at low SINR, as shown in Section \ref{sec:numerical}. Therefore, we will mainly focus on the joint optimization under low SINR.

	\section{PROBLEM FORMULATION AND PROPOSED JOINT OPTIMIZATION METHODS}

	In this letter, we aim to maximize the minimum rate of all users denoted as $R_\mathrm{sum} = \sum_{k=1}^{K}\log_{2}\left(1+\gamma_k\right) $ by jointly optimizing the active beamforming at the BS, passive beamforming at the IRS and dynamic PA.  The optimization problem is formulated as
	\begin{equation}
	\label{SUM SE PROBLEM}
	\begin{aligned}
	\begin{array}{ccl}
	& \mathop {\max }\limits_{\Theta,\{{p_k}\},\mathbf{F}} &\sum\limits_{k = 1}^K {{\log_{2}\left({1+\gamma_k}\right)}}  \hfill \\
	&\mathrm{s.t.} &{C_1:} \sum\limits_{k = 1}^K {{p_k}}\|\mathbf{F}\|_\mathrm{F}^2 \leqslant P, {p_k} \geqslant 0 ,\forall k, \hfill \\
	&&{C_2:}{R_k} \geqslant {R_{\min }}, \hfill \\
	&&{C_3:}|\theta_i| = 1, i = 1,2,\cdots,N_\mathrm{r},
	\end{array}
	\end{aligned}
	\end{equation}
	where each user is constrained by $C_1$ and $C_2$ to ensure that has a positive transmit power under data rate constraints, where $R_\mathrm{min}$ is the minimum data rate for each user. $C_3$ constrained the modulus constraint for IRS. It can be observed that every optimization variable entangle with each other in the form of  $\|\mathbf{h}_{k}\Theta\mathbf{H}\mathbf{F}\|_\mathrm{F}^2 p_k$ with quite complicated objective function and constraint $C_2$. Due to the problem \eqref{SUM SE PROBLEM} is non-convex, it is very difficult to obtain the closed-form solution to each variable.
	
	To overcome the difficulties mentioned above, we add relaxation variables to properly reconstruct the objective function into a linear form and obtain the suboptimal solution by using block alternating optimization. Specifically, the joint optimization problem can be decoupled into multiple independent optimization sub-problems, i.e., alternately optimize the transmission power, the full-digital precoding at the BS, and the optimal non-convex phase shift matrix with element-modulus constraints at the IRS. In addition, the non-convex joint optimization problem \eqref{SUM SE PROBLEM} is transformed as a quadratic programming function, by the extension of the Sherman Morrison-Woodbury formula \cite{MDC1999MJ}.
	
	Thus, the achievable rate maximization objective function is equivalent to minimum mean square error (MSE) structure, i.e., $\mathop {\max }\limits_{{p_k},\mathbf{\Theta},\mathbf{F}} \sum\limits_{k = 1}^K \mathop {\max }\limits_{q_k>0} {\left(\log _{2} q_k-\frac{q_k e_k}{\ln 2}\right)}$, where $q_k = \frac{1}{{e_{k}}}$ denotes the positive real number. $e_k $ is the MSE and given by
	\begin{equation}
	\label{e_k}
	e_k = |1-\sqrt{p_k}\mathbf{h}_{k}\Theta\mathbf{H}\mathbf{F}|^2+\left\|\mathbf{h}_{k}\Theta\mathbf{H}\mathbf{F}\right\|_\mathrm{F}^2\sum\limits_{i=1}^{k-1}p_i+\sigma_k^2.
	\end{equation}
	
	 Although the nonconvex objective function is rewritten as convex linear form, another difficulty for solving \eqref{SUM SE PROBLEM} lies in the QoS constraints that is seriously coupled to each variable. Therefore, we relax the constraint $C_2$ and decouple it into the optimization process of solving the full-digital precoding matrix $\mathbf{F}$ in the BS. Thus, for any given the reflection matrix $\Theta$ at the IRS and transmission power $p_k$, and introducing auxiliary variables $\{\mathbf{W},\hat{\mathbf{W}}\}$, the optimization problem for precoding matrix $\mathbf{F}$ is given by
		\begin{equation}
		\label{rewrite precoding F problem}
		\begin{aligned}
		\begin{array}{ccl}
		&\mathop {\min }\limits_{\mathbf{F},\mathbf{W},\hat{\mathbf{W}}} & C(\mathbf{F},\mathbf{W},\hat{\mathbf{W}}) \hfill \\
		&\mathrm{s.t.} &{C_4:}\sum\limits_{k = 1}^K {{p_k}}\|\mathbf{W}\|_\mathrm{F}^2 \leqslant P, {p_k} \geqslant 0 ,\forall k, \hfill \\ 		&&{C_5:}|\sqrt{q_k}-\hat{\mathbf{W}}|_\mathrm{F}^2+\left\|\hat{\mathbf{W}}\right\|_\mathrm{F}^2\sum\limits_{i=1}^{k-1}p_i \leqslant \varepsilon,\hfill \\
		&&{C_6:} \mathbf{W} = \mathbf{F},\\
		&&{C_7:} \sqrt{q_k p_k}\mathbf{h}_{k}\Theta\mathbf{H}\mathbf{F} = \hat{\mathbf{W}}.
		\end{array}
		\end{aligned}
		\end{equation}
		where $C(\mathbf{F},\mathbf{W},\hat{\mathbf{W}}) = \sum_{k = 1}^K ( |\sqrt{q_k}-\sqrt{q_k p_k}\mathbf{h}_{k}\Theta\mathbf{H}\mathbf{F}|^2+\left\|\sqrt{q_k p_k}\mathbf{h}_{k}\Theta\mathbf{H}\mathbf{F}\right\|_\mathrm{F}^2\sum_{i=1}^{k-1}p_i)$ and $\varepsilon = {\ln 2}(\log _{2}{q_k} - {R_{\min }})$. Note that, the problem is transformed as a multi-variable joint optimization problem with complete decoupling by define $\mathbf{W} = \mathbf{F}$ and $\hat{\mathbf{W}}=\sqrt{q_k p_k}\mathbf{h}_{k}\Theta\mathbf{H}\mathbf{F}$. Hence, problem \eqref{rewrite precoding F problem} can be solved  by further decoupling  and iterative optimization of two blocks, i.e., the original optimization variable problem $\mathcal{P}_1:\mathop {\min }\limits_{\mathbf{F}}  C(\mathbf{F},\mathbf{W},\hat{\mathbf{W}})$ with constraint $C_4$ and the equivalent variables problem $\mathcal{P}_2:\mathop {\min }\limits_{\hat{\mathbf{W},\mathbf{W}}}  C(\mathbf{F},\mathbf{W},\hat{\mathbf{W}})$ with constraint $C_5$. The augmented Lagrangian function of problem $\mathcal{P}_1$ and $\mathcal{P}_2$ are given by
		\begin{equation}
		\begin{aligned}
		&\mathfrak{L}_{\mathcal{P}_1} = C+ \mathbbm{1}_{C_4}(\mathbf{W}) + \frac{\varpi_{\mathcal{P}_1}}{2}\|\mathbf{F} + \frac{\lambda_{\mathcal{P}_1}}{\varpi_{\mathcal{P}_1}}- \mathbf{W}\|_\mathrm{F}^2 \\
		&\mathfrak{L}_{\mathcal{P}_2} = C +\mathbbm{1}_{C_5}(\hat{\mathbf{W}})+ \frac{\varpi_{\mathcal{P}_2}}{2}\|\sqrt{q_k p_k}\mathbf{h}_{k}\Theta\mathbf{H}\mathbf{F} + \frac{\lambda_{\mathcal{P}_2}}{\varpi_{\mathcal{P}_2}}- \hat{\mathbf{W}}\|_\mathrm{F}^2,
		\end{aligned}
		\end{equation}
		where  $\lambda_i$ and $\varpi_i$ $i \in \{\mathcal{P}_1,\mathcal{P}_1\}$ are the  Lagrange multiplier matrix and the scalar penalty parameter, respectively. $\mathbbm{1}_\mathcal{A}(W) = \left\{\begin{array}{ll}0, &w \in \mathcal{A} \\ \infty, &w \notin \mathcal{A}\end{array}\right.$. Therefore, problem \eqref{rewrite precoding F problem} can be addressed via the application of the ADMM \cite{IRSCE2020BZ} which involves the two-layer alternating minimization steps, the inner layer steps given by
		\begin{equation}
		\label{ADMM STEP1}
		\begin{aligned}
		&\mathbf{W}^t = \mathop{\mathrm{argmin}}\limits_{\mathbf{W}}\mathfrak{L}_{\mathcal{P}_1}\left(\hat{\mathbf{F}}^{t-1},\hat{\mathbf{W}}^{t-1},\mathbf{W},\lambda_{\mathcal{P}_2}^{t-1}\right) ,\\
		&\hat{\mathbf{W}}^t =  \mathop{\mathrm{argmin}}\limits_{\hat{\mathbf{W}}}\mathfrak{L}_{\mathcal{P}_2}\left(\hat{\mathbf{F}}^{t-1},\hat{\mathbf{W}},\mathbf{W}^{t-1},\lambda_{\mathcal{P}_2}^{t-1}\right),\\
		&\lambda_{\mathcal{P}_2}^{t} = \lambda_{\mathcal{P}_2}^{t-1} + \varpi_{\mathcal{P}_2}(\sqrt{q_k p_k}\mathbf{h}_{k}\Theta\mathbf{H}\mathbf{F}^{t-1} - \hat{\mathbf{W}}^{t}),
		\end{aligned}
		\end{equation}
		where $t$ is the iteration index. It is not difficult to observe that for any given $\mathbf{F}^{t-1}$, the constraint $C_5$ can still be satisfied by solving \eqref{ADMM STEP1} based on Lagrange duality method \cite{SCO2004} due to the addition of indicator function. Then, the outer layer update the $\mathbf{F}$ and the alternating steps are given by
		\begin{equation}
		\label{ADMM STEP2}
		\begin{aligned}
		&\mathbf{F}^t = \mathop{\mathrm{argmin}}\limits_{\mathbf{F}}\mathfrak{L}_{\mathcal{P}_1}\left(\hat{\mathbf{F}},\hat{\mathbf{W}}^*,\mathbf{W}^*,\lambda_{\mathcal{P}_1}^{t-1}\right)\\
		&\lambda_{\mathcal{P}_1}^{t} = \lambda_{\mathcal{P}_1}^{t-1} + \varpi_{\mathcal{P}_1}(\mathbf{F} - \mathbf{W}^*)
		\end{aligned}
		\end{equation}
		where $(\cdot)^*$ denote the optimal variable and the constraint $C_4$ can be ensured by solve $\mathbf{W}^t$. The termination criteria is given by $\|\mathbf{F}^t-\mathbf{F}^{t-1}\|_\mathrm{F}^2 \leq 10^{-3} \& \|\mathbf{F}^t- \mathbf{W}^*\|_\mathrm{F}^2 \leq 10^{-3}$.
		
		Moreover, the $q_k$ can be obtained by the first-order optimality condition of $e_k$  and independent of solving problem \eqref{rewrite precoding F problem}, which the optimal solution is given as
		\begin{equation}
		\label{q_k}
		q_k^t = \frac{1}{1-\sqrt{p_k}\mathbf{h}_{k}\Theta\mathbf{H}\mathbf{F}^t}.
		\end{equation}
		Accordingly, the high quality approximate solutions of $\mathbf{F}$ can be obtained by iteration \eqref{ADMM STEP1}, \eqref{ADMM STEP2} and \eqref{q_k}.  For the convergence of the above-relaxed method, the objective value of $\mathbf{F}$ achieved by the proposed ADMM algorithm is not increased over the two-step iteration, and the lower bound of \eqref{rewrite precoding F problem} is guaranteed by the transformed constraint $C_6$.  

	Next, for fixed precoding matrix $\mathbf{F}$ and any power factor $p_k$, the optimal phase shifts matrix is easily tackled by resorting the gradient-projection  method \cite{HPD2020CP} or the SCA method \cite{2020arXiv200500269Y}. Since QoS constraints are guaranteed in the precoding design at the BS, the optimization of the passive phase shifts matrix is limited by relaxed constraints, i.e., $\mathcal{S}_{C5}\{\forall \Theta^t|\sqrt{q_k}-\sqrt{q_k p_k}\mathbf{h}_{k}\Theta^t\mathbf{H}\mathbf{F}^*|_\mathrm{F}^2+\left\|\sqrt{q_k p_k}\mathbf{h}_{k}\Theta^t\mathbf{H}\mathbf{F}^*\right\|_\mathrm{F}^2\sum\limits_{i=1}^{k-1}p_i \leq \varepsilon\}$. Thus, the optimal $\Theta$ is given by
	
	\begin{equation}
	\label{hattheta}
	\hat{\Theta}^t = \Pi_{\mathcal{S}_{C_5}}\{\Theta^{t-1} - \mu \nabla\mathbf{\Theta}^{{(t-1)}}\}
	\end{equation}
	where $\nabla\mathbf{\Theta}^{{(t-1)}} =\mathbf{h}_k^\mathrm{H}((p_k+\sum_{i=1}^{k-1}p_i)\mathbf{h}_k\mathbf{\Theta}^{(t-1)}\mathbf{H}\mathbf{F}^*-2\sqrt{p_k})\mathbf{F}^{*\mathrm{H}}\mathbf{H}^\mathrm{H}$, denote the first-order gradient of the previous iteration and $\mu$ is a step size parameter. $\Pi_{\mathcal{S}_{C_5}}$ is the projection onto the set $\mathcal{S}_{C_5}$ and is solved base on  a similar calculation process in \cite{HPD2020CP}. Note that, an additional calculation process needs to be considered due to the unique modulus constraints $C_3$, that is
	\begin{equation}
	\label{theta}
	\Theta^t = \mathrm{diag}(e^{\mathcal{J} \angle \hat{\Theta}^t}) .	
	\end{equation}
	
	Through the above iterative solution, the objective function can be simplified to solve for variable $p_k^t$. For simplicity, transforming QoS constraints $C_2$ to the convex linear form $C_{8}:\left\|\mathbf{h}_k\Theta\mathbf{H}\mathbf{F}\right\|_\mathrm{F}^{2} p_{k}-(2^{\mathbf{R}_\mathrm{min}}-1)\left\|\mathbf{h}_k\Theta\mathbf{H}\mathbf{F}\right\|_\mathrm{F}^{2} \sum_{i=1}^{k-1} p_{i} \geq (2^{\mathbf{R}_\mathrm{min}}-1) \sigma_{k}^{2}$, and the optimization problem for $p_k$ at  $t$th Alternate optimization process can be rewritten as
		\begin{equation}
		\label{rewrite p_k problem}
		\begin{aligned}
		\begin{array}{ccl}
		&\mathop {\max }\limits_{\{p_k^t\}} &\sum\limits_{k = 1}^K q_k^t(|1-\sqrt{p_k^t}\mathbf{h}_k\Theta^t\mathbf{H}\mathbf{F}^t|^2+\varkappa_k) \\&&+ \mathbbm{1}_{C_8}(p_k^t) +\upsilon\left(\sum_{k=1}^{K}p_i\|\mathbf{h}_k\Theta^t\mathbf{H}\mathbf{F}^t\|_\mathrm{F}^2-P\right)  \hfill \\
		&\mathrm{s.t.} & \upsilon \ge 0
		\end{array}
		\end{aligned}
		\end{equation}
		where $\upsilon$ denote the Lagrange operator and $\mathbbm{1}_{C_8}(p_k^t)$ is the indicator function which defined as mentioned above.
		
		It can be seen that the objective function of \eqref{rewrite p_k problem} is a convex optimization problem with no constraint which can be efficiently solved by the Karush-Kuhn-Tucker (KKT) condition. The analytic solution for optimization objective $p_k^t$ can be obtained from the first-order gradient, which is given by
		\begin{equation}
		\nabla p_k^t = \left(\frac{q_k^t\mathrm{Re}(\mathbf{h}_k\Theta^t\mathbf{H}\mathbf{F}^t)}{\varsigma}\right)^2,
		\end{equation}
		where $\varsigma=q_k^t\left\|\mathbf{h}_k\Theta^t\mathbf{H}\mathbf{F}^t\right\|_\mathrm{F}^2+\sum_{i=k}^{K}\left\|\mathbf{h}_k\Theta^t\mathbf{H}\mathbf{F}^t\right\|_\mathrm{F}^2q_i+\upsilon$.  
	
	\begin{algorithm}[!t]	
		\SetAlgoNoLine 
		\caption{The Iterative Alogrithm for Joint Optimization}
		\label{alg:power}
		\LinesNumbered
		\KwIn{$P,T_\mathrm{max},\mathbf{H},\mathbf{h}_k$}
		\KwOut{$\mathbf{F},\Theta,p_k$}
		\textbf{Initialize}
		Initial power for each user $p_k^0$,  the initial matrix $\Theta^0$ and the auxiliary variable $q_k^0$, \\
		$ t=0$.\\
		\While{$t\le T_\mathrm{max}$}{
			Compute optimal $\mathbf{F}^t$ from \eqref{ADMM STEP1} and \eqref{ADMM STEP2};\\
			update  $q_k^t$  using \eqref{q_k};\\
			Compute optimal $\Theta^t$ from \eqref{theta};\\
			Compute $p_k^t$ by solving the equation of \eqref{rewrite p_k problem}; \\
			$t= t+1$.\\
		}
	\end{algorithm}
	
	It can be easily verified that each optimal intermediate variables can be obtained in the $T_\mathrm{max}$th iteration due to the limitation of the maximum transmission power $P$ and the characteristics of each objective function. To find the stationary solution for each variable, the overall algorithm framework is summarized in \textbf{Algorithm \ref{alg:power}}  based on successively updating  three  optimization block $\mathbf{F}$, $\Theta$ and $q_k$ via \eqref{rewrite precoding F problem}, \eqref{theta} and \eqref{rewrite p_k problem}, where $p_k^0$ and $\Theta^0$ are initial transmit power and reflecting phase shifter matrix, respectively.
	
	Firstly, the precoding design in BS is optimized by double-layer ADMM iteration, which mainly involves the matrix-vector multiplication with complexity $O(N_t^3+K)$ in outer layer and $O(K^3)$ in inner layer. Thus, the total complexity of optimized the precoding matrix $\mathbf{F}$ is  $O(K^3N_\mathrm{t}^3+K^4)$. Next, the gradient projection algorithm is used to obtain the suboptimal solution of the reflection phase-shift matrix $\Theta$, and the complexity is $O(3N_\mathrm{t} N_\mathrm{r}^2+N_\mathrm{r}^3)$. For the dynamic PA, the main complexity originates comes from solving the convex optimization problem \eqref{rewrite p_k problem} in the $T_\mathrm{max}$th iteration, where the complexity is related to the number of users, i.e., $O(T_\mathrm{max}K)$. Hence, the total computational complexity of solving the joint optimization problem \eqref{SUM SE PROBLEM}  is $O(T_\mathrm{max}K+ K^3N_\mathrm{t}^3+K^4 + 3N_\mathrm{t} N_\mathrm{r}^2+N_\mathrm{r}^3)$.

	\section{NUMERICAL RESULTS}
	\label{sec:numerical}
	In this section, an IRS-aided system is considered for the simulation, where the BS equipped with N = 256 antennas. Moreover, considering the channel parameters of user $k$, among them $L_\mathrm{IU}= L_\mathrm{BI} = 3$, $\mathfrak{p}^{\mathrm{BI}}_{n}$ and $\mathfrak{p}_{k, n}^{\mathrm{IU}}$ follow $\mathcal{C}\mathcal{N}(0,10^{\mathcal{P}(d_\mathrm{bu})})$ \cite{MWCM2014AMR}, where $\mathcal{P}(d_\mathrm{bu}) = 61.4\mathrm{dB}+20\log_{10}(d_\mathrm{bu})+\xi, \xi\sim\mathcal{C}\mathcal{N}(0,\sigma_{\xi}^{2})$, with $d_\mathrm{bu}$ as the width of IRS. The minimal achievable rate for each user $\mathbf{R}_\mathrm{min}$ is assumed  as $0.01\mathrm{bit/s/Hz}$. It is assumed that each user has the same noise power $\sigma_k$ and the signal-to-noise ratio (SNR) is defined as $10\log_{10}(\frac{P}{\sigma_{k}})$.
	
	\begin{figure}[!ht]
		\centering
		\includegraphics[width=0.6\linewidth]{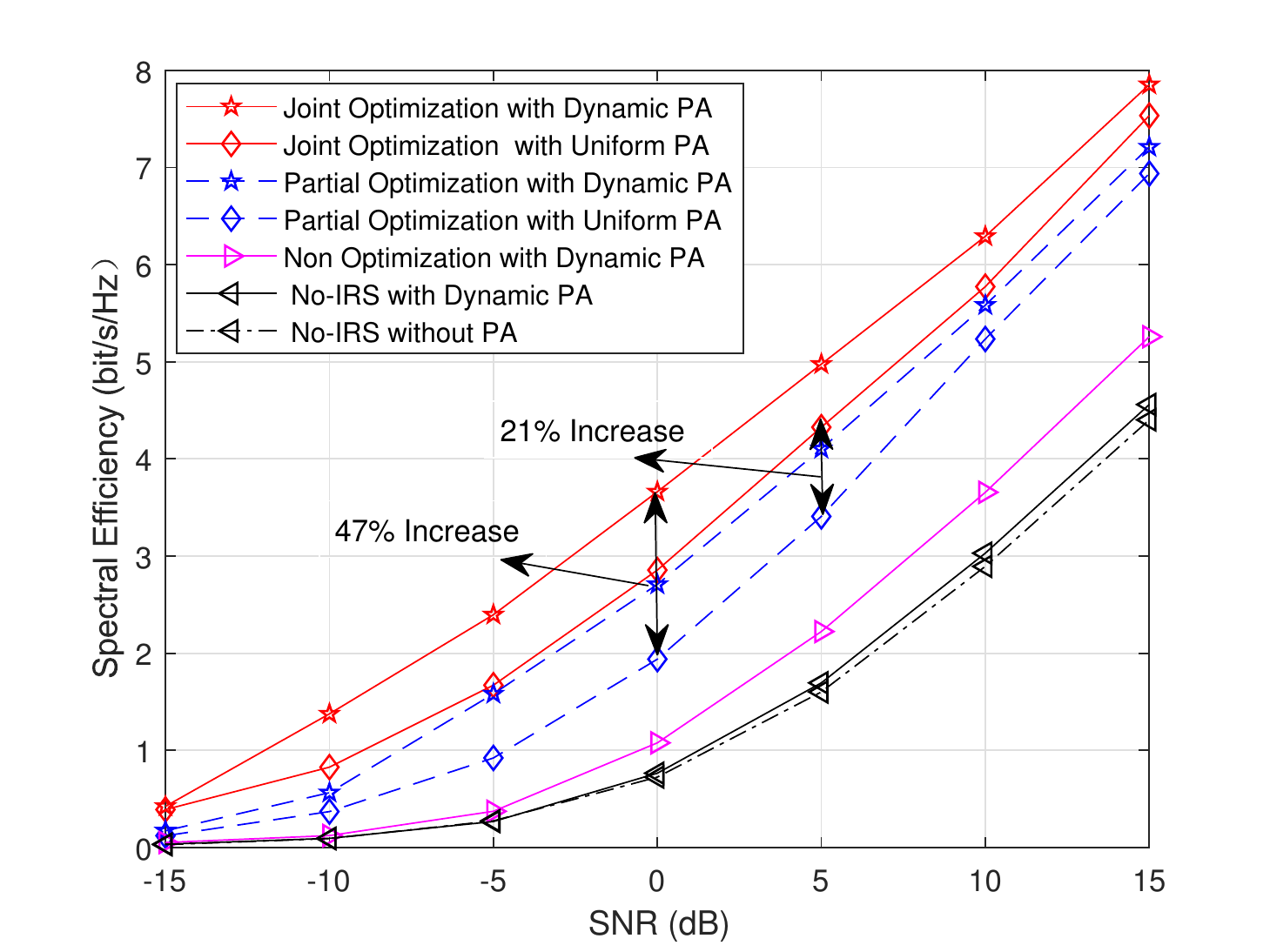}
		\caption{\footnotesize  Spectrum efficiency versus the increase of SNR, where $P = 10 \mathrm{dBm}$, $K = 16$, $N_\mathrm{r} = 64$.}
		\label{SE-snr}
	\end{figure}
	Fig. \ref{SE-snr} shows the sum-SE of four structures, where in partial optimization scenario, the precoder $\mathbf{F}$ does not participate in the joint optimization and is fixed as the channel space vector. Similarly, in the non-optimization scenario, the IRS reflection matrix with random element values is not involved in the joint optimization. Moreover, the system without IRS-assisted is regarded as the lower bound for comparison. Comparing to the partial optimization scenario, more than $47\%$ improvement of SE based on the proposed joint optimization algorithms. However, if the dynamic PA does not participate in the joint optimization, only $21\%$ improvement can be achieved based on the uniform PA. Therefore, in order to obtain satisfactory SE performance, the proposed dynamic PA scheme should participate in the joint optimization. 
	
		\begin{figure}[!ht]
		\centering
		\includegraphics[width=0.6\linewidth]{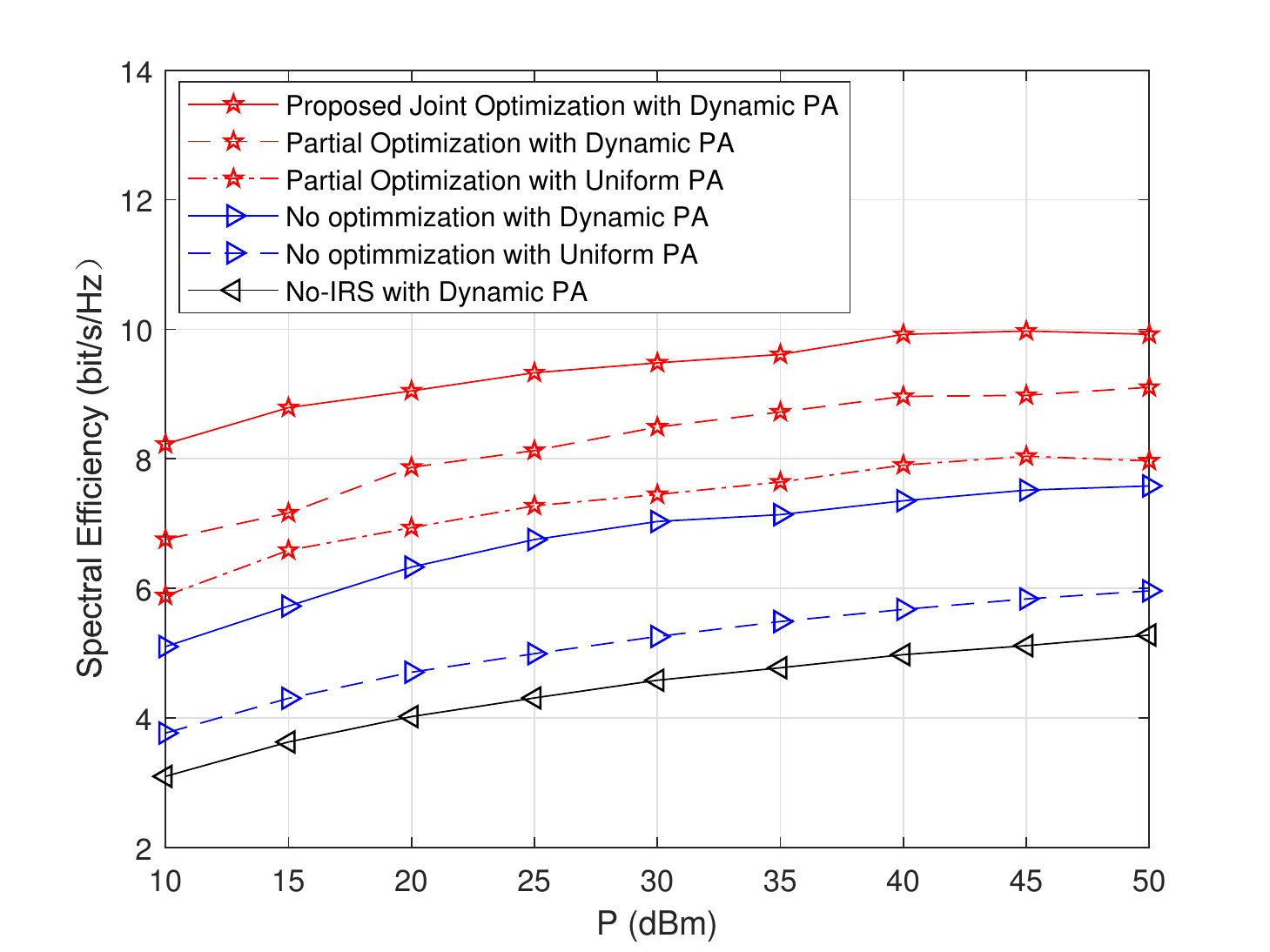}
		\caption{\footnotesize  Spectrum efficiency versus the maximum transmit power $P$, where $\mathrm{SNR} = 10 \mathrm{dB}$, $K = 16$, $N_\mathrm{r} = 64$.}
		\label{SE-p}
	\end{figure}
	
	Similarly, Fig. \ref{SE-p} shows the SE where $P$ varies from $10 \mathrm{dBm}$ to $50 \mathrm{dBm}$ with the different optimization methods.  It can be seen that as the limit conditions of the total power constraint $P$ is relaxed (the number of users is fixed but the total power is increased), the SE performance is based on each optimization scheme is slowly increasing. Among them, the proposed joint optimization scheme still has excellent performance.

	\begin{figure}[!ht]
		\centering
		\includegraphics[width=0.6\linewidth]{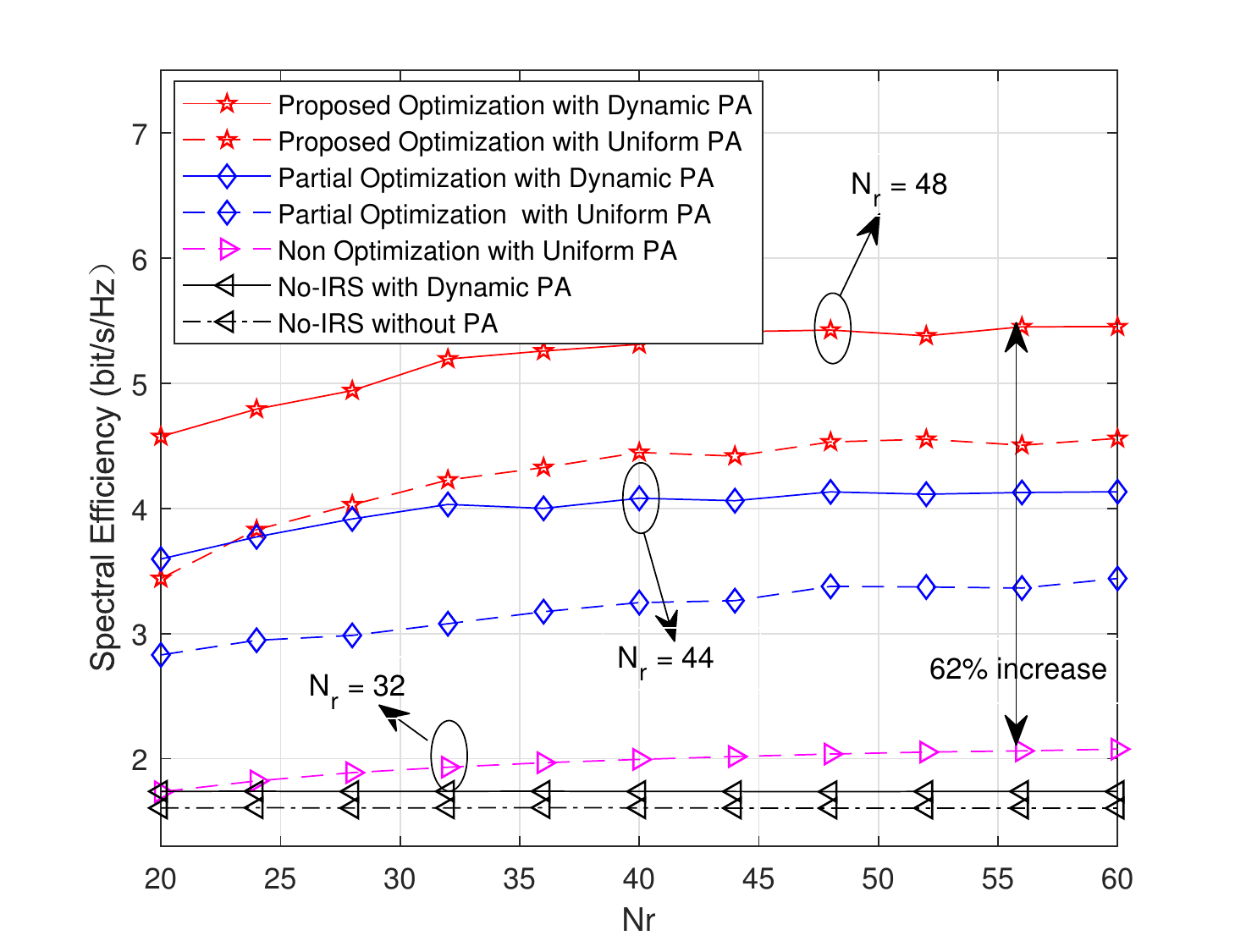}
		\caption{\footnotesize  Spectrum efficiency versus the increase of $N_\mathrm{r}$, where $P = 5 \mathrm{dBm}$, $\mathrm{SNR} = 5\mathrm{dB}$, $K = 16$.}
		\label{SE-nr}
	\end{figure}
	
	Fig. \ref{SE-nr}  shows the sum-SE by different approaches with the increasing IRS elements. It is observed that as $N_\mathrm{r}$ increases, the gradient of achievable SE decrease gradually due to the limitation of the maximum transmission power at the BS. Specifically, the proposed joint optimization algorithm has higher unit adaptability which the upper bound is $N_\mathrm{r} = 48 $. It is noteworthy that the SE based on the proposed algorithm has greatly improved than the schemes which adopted the dynamic PA but without optimization. Besides, since the impact of the modulus constraints  $C_3$ at the IRS has been compensated by the joint optimization of the precoding and reflection matrix, it is possible to further promote the SE by applying the proposed dynamic PA even if the system utilizes large-scale IRS units.
	
	\section{CONCLUSION}
	This letter investigated the downlink of IRS-assisted massive MIMO systems with multi-user interference and transmission power limit. A dynamic joint optimization scheme for maximizing the sum-SE was proposed with iterative optimization under the QoS constraints of all users. In the derivations, new methods for complex coupling variables were developed in restructured the multivariable optimization object with the assistance of the transformational relation between the MSE and achievable rate.  Besides, the suboptimal solution of height approximation for each iteration variable was provided.Numerical results validated the superiority of the proposed joint optimization algorithm. Moreover, it was shown that the sensitivity of the performance upper bound improved additionally to the reflection units, illustrating the effectiveness of employing the joint optimization with dynamic PA  at the IRS-assisted system.

	\bibliographystyle{IEEEtran}
	\bibliography{bare_jrnl}
	
\end{document}